\documentclass[twocolumn,prl,superscriptaddress,amsmath,amssymb,mathrsfs,showpacs]{revtex4-2}
\usepackage{tabularx}
\usepackage{graphicx}
\usepackage{braket}
\usepackage[utf8]{inputenc}

\usepackage{dcolumn}
\usepackage{bm}
\usepackage{epsfig}
\usepackage{rotating}
\usepackage{color}
\usepackage[colorlinks,citecolor=blue]{hyperref}

\newcommand{\change}[1]{\textcolor{black}{#1}}

\begin{document}

\title{Sympathetic rotational cooling of large trapped molecular ions}
\author{Monika Leibscher}
\affiliation{Freie Universit\"{a}t Berlin, Dahlem Center for Complex Quantum Systems and Fachbereich Physik, Arnimallee 14, 14195 Berlin, Germany}
\author{Alexander Blech}
\affiliation{Freie Universit\"{a}t Berlin, Dahlem Center for Complex Quantum Systems and Fachbereich Physik, Arnimallee 14, 14195 Berlin, Germany}
\author{Christiane P. Koch} 
\email{Email: christiane.koch@fu-berlin.de}
\affiliation{Freie Universit\"{a}t Berlin, Dahlem Center for Complex Quantum Systems and Fachbereich Physik, Arnimallee 14, 14195 Berlin, Germany}
\date{\today} 

\begin{abstract}
  We suggest a protocol for the sympathetic cooling of a molecular asymmetric top rotor co-trapped with laser-cooled atomic ions, 
  based on resonant coupling between the molecular ion's electric dipole moment and a common \change{normal} mode of the trapped particles. 
  By combining sympathetic sideband laser cooling with coherent microwave excitation, we demonstrate the efficient depopulation of arbitrary rotational subspaces and the ability to cool an incoherent distribution of rotational states into a single, well-defined quantum state. This capability opens the door to exploiting the rotational Hilbert space for applications in quantum information processing and high-precision spectroscopy.
 \end{abstract}

\maketitle

Polyatomic molecules are emerging as a \change{versatile} platform for fundamental physics \cite{SafronovaRMP18,HutzlerQST20} and quantum information processing \cite{AlbertPRX20}. 
\change{They offer, for instance, high sensitivity in searches for the electron's} dipole moment~\cite{AndereggScience2023} or parity violation in chiral \change{species}~\cite{QuackChemSci2022,ErezPRX23}. Molecular rotation, \change{owing to its large, well-isolated Hilbert space,} can be used  for quantum error correction~\cite{AlbertPRX20,Shubham_2023,FureyQuantum24} and quantum simulation of magnetism~\cite{WallAnnPhys2013}. 
Linear rotors and asymmetric top rotors are completely controllable within finite subspaces~\cite{Judson1990, Pozzoli21, Leibscher22}, and the anharmonicity of the rotational spectrum facilitates quantum state control~\cite{KochRMP19}. Consequently, any desired unitary operation on the subspace can be implemented with a combination of resonant electric fields --- typically in the microwave range --- notwithstanding spectral degeneracies~\cite{Leibscher22}.
However, a prerequisite to implementing the desired level of control is the trapping and cooling of molecules~\cite{Deiss2023,Sin22b}. 

\change{The charge of molecular ions facilitates trapping with radio-frequency fields~\cite{Deiss2023,Sin22b}, and the Coulomb interaction enables indirect control of the molecules via co-trapped atomic ions. Quantum logic spectroscopy represents a landmark implementation of this approach, allowing for state-selective preparation and non-destructive readout of diatomic molecules~\cite{WolfNature_2016,ChouNature2017,ChouSci202,SinhalScience2020}. 
The Coulomb interaction also facilitates sympathetic cooling of molecular translational motion, even for complex polyatomic species~\cite{Ostendorf_2006,HojbjerrePRA2008,SchmidJPCA2022,Calvin_Nature_2023,Xu_2024}, 
using laser-cooled atomic ions. 
However, internal state preparation for polyatomic molecules remains challenging~\cite{Petterson_2018}. Their considerably larger state space, relative to diatomic molecules, hinders the effectiveness of both quantum logic spectroscopy~\cite{WolfNature_2016,ChouNature2017,ChouSci202,SinhalScience2020} and optical cooling methods~\cite{Staanum_2010,Schneider_2010,Lien_2014}. Collisions with a buffer gas allow for cooling molecular vibrations~\cite{GerlichFaraday2009,SchmidJPCA2022,CalvinPRA2023} and, in case of diatomics, also rotations~\cite{Hansen_2014,HolzapfelPRX2025}. For polyatomic molecules, however, cooling the rotational degrees of freedom is an open challenge. 
}

\begin{figure*}[tbp]
    \includegraphics[width=\linewidth]{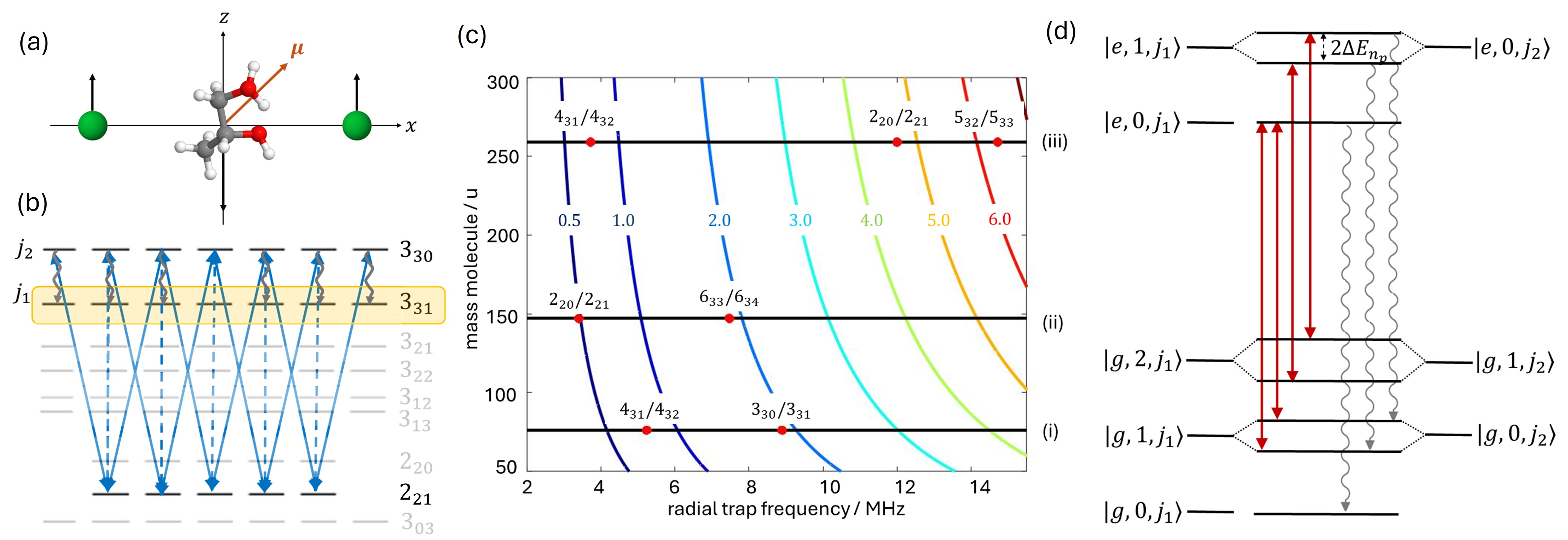}  
	\caption{\change{(a) Trapping a polyatomic ion with intrinsic dipole moment $\bm \mu$ (here protonated 1,2-propanediol) together with two Yb$^+$ ions. The black arrows indicate the radial zig-zag normal mode. 
    (b) Sympathetic rotational cooling of the molecule, modeled as an asymmetric top rotor, combines an effective decay of the $j_2$-level (wiggled gray arrows) with resonant microwave excitation depleting all other rotational states. Solid (dashed) blue arrows indicate transitions induced by $x$- and $z$-polarized pulses with selection rules $\Delta M= \pm 1$ and $\Delta M = 0$, respectively.
    The effective decay is due to sideband-cooling of the atoms and requires resonant dipole-phonon coupling. 
    (c) Dipole-phonon coupling strength (see also SM) in kHz/D (contour lines) vs radial trap frequency $\omega_z$ and molecular mass (with axial trap frequency $\omega_x=1$ MHz). The horizontal lines indicate the masses of (i) protonated 1,2-propanediol, (ii) protonated glutamine (ii), and (iii) CHCaBrI$^+$; red stars point to dipole-allowed transitions between asymmetric top eigenstates $J_{K_a,K_c}$ and $J_{K_a,K_{c+1}}$ which are resonant to the frequency $\omega_p$ of the radial zig-zag mode. 
    (d) Combined level scheme with $\ket{a,n_p,j} = \ket{a} \otimes \ket{n_p,j}$, where $a=g,e$ denotes the internal state of the atom (for simplicity, only the levels of one atom are depicted), $n_p$ the phonon excitation, and $j_{1/2}$ the rotational state. Dipole-phonon coupling induces a splitting $2 \Delta E_{n_p}$ of the states $\ket{a,n_p,j_1}$ and $\ket{a,n_p-1,j_2}$, cf. Eq.~(\ref{eq:dipole_splitting}). The wiggled gray arrows indicate spontaneous emission on the atom, the red arrows show the transitions induced by the cooling laser.}
    }
	\label{fig:trap}
\end{figure*}
\change{Here we propose to sympathetically cool the rotational motion of polyatomic molecules which are co-trapped with atomic ions. The protocol is based on two key components, illustrated in Fig.~\ref{fig:trap}: 
engineering dissipation on one rotational transition by resonantly coupling it to the sympathetically cooled trap motion, and using coherent microwave excitation to pump population from all other rotational levels into the cooled transition. Suitable microwave drives can always be found due to the complete controllability of asymmetric top rotors~\cite{Leibscher22,Pozzoli21}. The required coupling between molecular rotation and translational motion is provided by the intrinsic dipole moment of polar molecules; this interaction becomes strong when a rotational transition matches a trap normal mode frequency~\cite{HudsonPRA2018,Campbell_PRL_2020,Qi_2024_arxiv}.
We show that rotational transitions enabling resonant coupling with the trap motion can be found across a large variety of molecular species and predict rotational cooling times on the order of milliseconds.}

We consider a polyatomic molecular ion co-trapped in a linear Paul trap with two atomic ions, cf. Fig.~\ref{fig:trap}(a), 
\change{ and assume that molecular vibrations have been cooled to the ground state  using e.g. a buffer gas~\cite{GerlichFaraday2009,SchmidJPCA2022,CalvinPRA2023}.}
\change{The translational motion of the particles in the trap} is described by normal modes oscillating with frequencies $\omega_p$ in a harmonic potential $V$. 
\change{ Sideband laser cooling of the atoms \cite{Diedrich_PRL_1989} sympathetically cools the molecular translational motion provided both}
atoms and molecule oscillate with non-vanishing amplitude about their equilibrium position. 
For polar molecules, the \change{intrinsic} molecular dipole moment \footnote{\change{ The notion of a "permanent" molecular dipole moment might be misleading, since P and T symmetry forbid the existence of a permanent dipole moment. The (intrinsic) molecular dipole moment arises from the mixing of quasi-degenerate states with opposite parities \cite{Tino_2022}.}} couples the molecular rotation to the \change{trap normal modes ("phonons")}, $H_{dp}=-{\bm \mu}  \cdot {\bf R}\cdot {\bf E}$~\cite{HudsonPRA2018,Campbell_PRL_2020,Leibscher_Kahler_Koch_arxiv}. 
\change{Here,} ${\bf R}$ is the rotation matrix describing the orientation of the molecule-fixed dipole moment ${\bm \mu}$ in the laboratory frame, \change{and ${\bf E} = - \nabla V$ is the electric field to which the dipole moment is exposed.}
If one of the trap normal mode frequencies $\omega_p$, \change{the radial zig-zag mode in our example, cf. Fig.~\ref{fig:trap}}, is resonant to a dipole-allowed transition between two rotational states, these two states are strongly mixed while dipole-phonon coupling with all other rotational states is negligible.
Denoting the bare states of the trapped particles by $|n_p,j\rangle = \ket{n_p} \otimes \ket{j}$, where $n_p=0,1,2,...$ counts the trap excitations and $\ket{j}$ is the rotational state of the molecule (for simplicity we omit the internal states of the atomic ions which are not affected by the dipole-phonon coupling), the interaction-dressed states are 
\begin{eqnarray}
     |n_p,j_\pm \rangle = \frac{1}{\sqrt{2}} \left( |n_p,j_1 \rangle \pm  |n_p-1,j_2 \rangle \right)\,.
\end{eqnarray}
\change{Their complete mixing is due to the degeneracy of the bare states $|n_p,j_1 \rangle$ and $|n_p-1,j_2 \rangle$~\cite{HudsonPRA2018,Campbell_PRL_2020,Leibscher_Kahler_Koch_arxiv}, and 
the dressed-state levels are split by
}
\begin{equation}
 \Delta E_{n_p} = |\bra{n_p,j_1} H_{dp} \ket{n_p-1,j_2}|,
 \label{eq:dipole_splitting}
\end{equation}
see Fig.~\ref{fig:trap}\change{(d)}.
\change{
While the detailed dependence on molecular parameters is discussed in the Supplemental Material (SM)~\cite{SM}, 
resonant coupling of a few kHz per 1$\,$D of dipole moment is attainable
for molecules with masses between 70 and 270$\,$u and radial trap frequencies below 15$\,$MHz. 
}
For $n_p>0$, the coupling between the states $\ket{n_p,j_1}$ and 
$\ket{n_p-1,j_2}$ \change{engineers an effective decay of} the rotational state
$\ket{j_2}$: Due to the coupling, laser cooling \change{ of the atoms} does not only cool \change{the phonons, i.e.,} the joint translational motion of atomic and molecular ions in the trap, but also the rotational population in level $j_2$ to level $j_1$. 
\change{One can think of it is as first- and second-order sympathetic cooling.}
All other rotational states are not sufficiently affected by the dipole-phonon interaction to be cooled, but their population can be \change{shuffled to $j_2$}  with microwave pulses. \change{Repeated cycles of cooling and microwave excitation then deplete all rotational excitations, as we demonstrate below for protonated 1,2-propanediol (and for protonated glumatine in the SM~\cite{SM}).}

\begin{figure*}[tbp]
\includegraphics[width=0.9\linewidth]{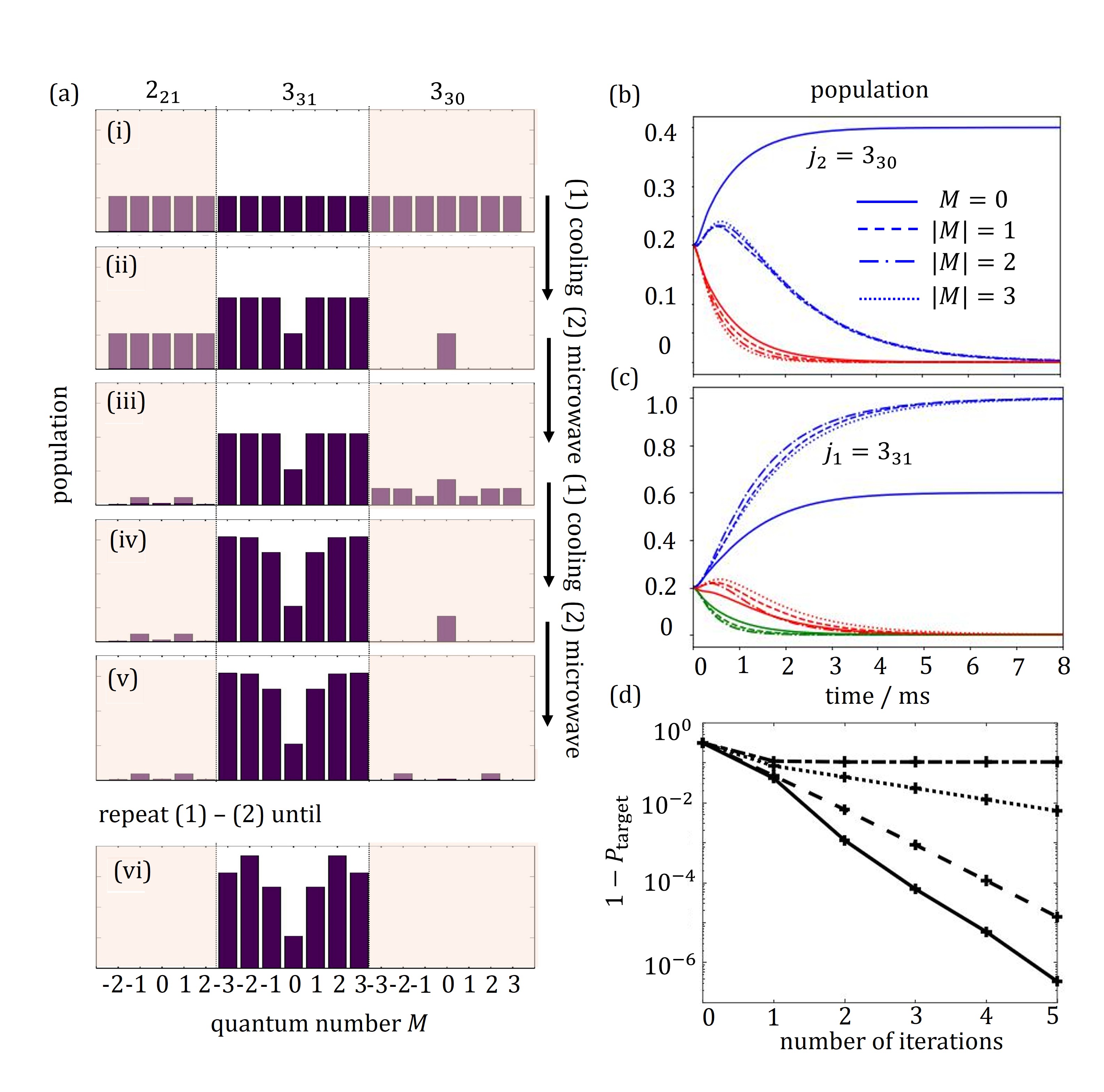}
   \caption{\change{Sympathetic cooling of protonated 1,2-propanediol into the rotational level $3_{31}$, cf. Fig.~\ref{fig:trap}(b), with repeated cycles of laser cooling and microwave population shuffling. (a) Total population of the states $\ket{2_{21},M}$, $\ket{3_{31},M}$ and $\ket{3_{30},M}$  with the atom in its ground state. 
    (b,c) $M$-resolved populations 
    for one step of laser cooling.
    The parameters for laser cooling are:} Rabi frequency $\Omega = 2 \pi \times 200\,$kHz, effective decay rate $\gamma = 0.1\,$MHz, Lamb-Dicke parameter $\eta = 0.012$ \cite{Kulosa_2023}, see also SM~\cite{SM}. 
    \change{The blue, red and green} curves correspond to $n_p=0,1,2$.
    (d) \change{Cooling error, } $1-\sum_M P_{3_{31},M}$, vs number of iterations for microwave pulses with perfect $x$-polarization (solid line), and with 25\% (dashed line), 50\% (dotted line), and 100\% (dash-dotted line) $z$-polarization.  
    }
	\label{fig:sideband_cooling}
\end{figure*}
\change{Due to their sequential application, sideband cooling and microwave population shuffling can be modeled separately. Sideband cooling is achieved by driving the atomic ions with laser pulses followed by rapid spontaneous emission. The laser frequency is assumed to be far off-resonant from any molecular transition, such that the molecules do not interact directly with the light. We describe}
the cooling 
by solving the master equation,
\begin{eqnarray}
    \frac{\partial \rho}{\partial t} = - \frac{i}{\hbar} \left[ H(t), \rho \right] +
    {\cal L}_D \rho\,,
    \label{eq:Lindblad}
\end{eqnarray}
where $H(t)$ describes the coherent dynamics induced by the laser, cf. Eq.~\eqref{eq:H_of_t}, and ${\cal L}_D$ the spontaneous emission on the atoms,
cf. End Matter. 
\change{Microwave excitation can be restricted to the rotational subspace and is} described by the Hamiltonian
\begin{equation}
H_{MW} = -\bm{\mu} \cdot {\bm R}\cdot \bm{E}_{MW} (t)\,,
\label{eq:H_MW}
\end{equation}
where $\bm{E}_{MW}(t) = \bm{e} {\cal E} (t) \cos(\omega t)$ is a linearly polarized  electric field with polarization direction $\bm{e}$, amplitude ${\cal E} (t)$ and frequency $\omega$.  
The Schrödinger equation for the rotational dynamics is solved as described in \cite{Leibscher22}. Dipole selection rules allow transitions with $\Delta J = 0, \pm 1$;
$z$-polarized fields induce transitions with $\Delta M=0$ (dashed blue arrows in Fig.~\ref{fig:trap}(c)), whereas $x$- and $y$-polarized fields drive transitions with $\Delta M = \pm 1$ (solid blue arrows in Fig.~\ref{fig:trap}). 

\begin{figure}[tbp]
	\includegraphics[width=0.8\linewidth]{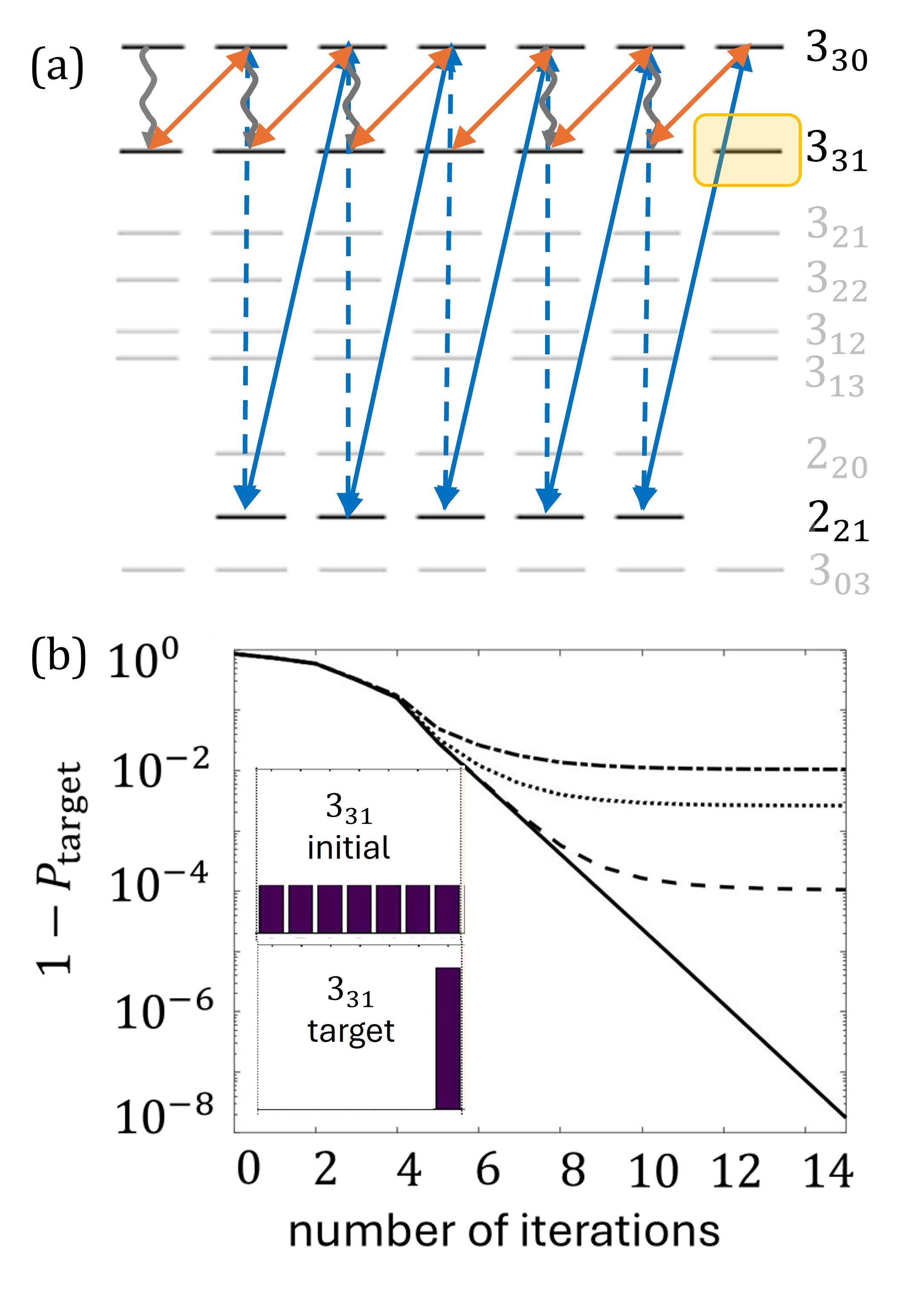}
	\caption{\change{Sympathetic rotational cooling to prepare a single quantum state of protonated 1,2-propanediol within a degenerate manifold:} (a) Microwave excitation (blue and orange arrows) to \change{prepare} the state $\ket{3_{31},3}$, marked by the yellow box, with the same \change{engineered decay from the states $\ket{3_{30},M}$ as in Fig.~\ref{fig:trap}(b).} (b) \change{Cooling error vs} number of iterations for microwave pulses with perfect $\sigma_+$-polarization (solid line) and pulses with 1\% (dashed line), 5\% (dash-dotted line), and 10\%  (dotted line) $\sigma_-$polarization. The inset shows the initial and target population \change{distributions}.}
	\label{fig:cooling_degenerate}
\end{figure}

\change{Taking the example of protonated 1,2-propanediol, resonance of the rotational transition $\ket{3_{30},M} \leftrightarrow\ket{3_{31},M}$ with the radial zig-zag mode occurs for a trap frequency of 8.84$\,$MHz.}
We consider a subset of the rotational states, as depicted by the dark gray lines in Fig.~\ref{fig:trap}(b),
\change{and} assume that initially all states of this rotational subsystem are populated, as depicted in panel (i) of Fig.~\ref{fig:sideband_cooling}(a). The target is to deplete all rotational states except $\ket{3_{31},M}$, see panel (vi) of Fig.~\ref{fig:sideband_cooling}, or, \change{more ambitiously,} to cool the rotational ensemble to a single quantum state, cf. Fig.~\ref{fig:cooling_degenerate}(a).
\change{When applying the sideband cooling step, it takes about 8$\,$ms to deplete the rotational states $3_{30}$, except for $M=0$ which is not affected by the cooling.
Although the coupling strength depends on the orientational quantum number, 
cf. End Matter,  
and the splittings vary from $\Delta E_{n_p,j} = 8.4 \sqrt{n_p}\,$kHz for $|M|=3$
to $2.7 \sqrt{n_p}\,$kHz  for $|M|=1$, the $M$-dependence is barely noticeable in Fig.~\ref{fig:sideband_cooling}(b,c).
The population distribution after the laser cooling step is shown in panel (ii) of Fig.~\ref{fig:sideband_cooling}(a).
Next, applying an $x$-polarized microwave pulse resonant with the $3_{31}\leftrightarrow 2_{21}$ transition removes population from both $\ket{3_{31},M=0}$ and the level $2_{21}$, cf. panel (iii) of Fig.~\ref{fig:sideband_cooling}(a).
Alternating sideband cooling and coherent microwave excitation accumulates the population in the target rotational level, see panels (iv) - (vi) of Fig.~\ref{fig:sideband_cooling}(a). The target is reached with very good accuracy after only a few iterations, cf. Fig.~\ref{fig:sideband_cooling}(d).}
In an experiment, the microwave pulses typically have a slight ellipticity instead of being perfectly linearly polarized but 
Fig.~\ref{fig:sideband_cooling}(d) shows that \change{the protocol achieves excellent cooling errors even with 25\% admixture of $z$-polarization and only 1 or 2 more iterations. Only}
with a purely z-polarized pulse (dash-dotted line), the target cannot be reached, \change{since without} an 
$x$- (or $y$-) component 
population \change{gets} trapped in $\ket{3_{30}, M=0}$.

\change{In order to cool a larger rotational state space, more drives -- one for each additional level, with frequency matching a transition involving this level -- need to be included in the microwave excitation step. A}
 quantum asymmetric top is controllable, 
\change{i.e., there exist pulse} sequences to deplete \change{the additional levels, provided the intrinsic molecular} dipole moment has at least two non-vanishing \change{Cartesian projections} \cite{Leibscher22}. \change{For levels that do not have a dipole-allowed transition to $j_2$, }
e.g. because $|\Delta J| > 1$, the population 
must first be \change{sequentially transferred to}
an already empty level \change{with a dipole-allowed transition to $j_2$}
before the \change{next cooling step. Eventually, all population accumulates in $j_1$, i.e., population in
arbitrary rotational subspaces can in principle be cooled}. 

While the protocol of Fig.~\ref{fig:sideband_cooling} allows for cooling arbitrary distributions of rotational states to the degenerate level $3_{31}$, it does not allow \change{for preparing} a single rotational quantum state. \change{This}
requires 
at least two different microwave frequencies for coherent population transfer, as depicted in Fig.~\ref{fig:cooling_degenerate}(a).
\change{Starting} with an incoherent ensemble in level $3_{31}$, the target state is 
$\ket{3_{31},3}$, marked in yellow in Fig.~\ref{fig:cooling_degenerate}(a). 
The \change{protocol now contains two stages of repeated cooling and microwave excitation, where the first stage uses} a circularly polarized pulse, \change{resonant} with $E_{3_{30}} - E_{3_{31}}$ \change{and} exciting $\Delta M = + 1$ (orange arrows in Fig.~\ref{fig:cooling_degenerate}(a)).
The combination of coherent $\Delta M = +1$ transitions and cooling along a $\Delta M = 0$ transition shifts the population toward the rotational state with the largest value of $M$, i.e., $M=3$. However, since there is no decay channel for $M=0$, some population remains trapped in $\ket{3_{30},0}$. This can be amended by \change{the second stage of the protocol,}
driving the population from $\ket{3_{30},0}$ {\it via} $\ket{2_{21},0}$  to
$\ket{3_{30},1}$, using a combination of $z$- and $\sigma_+$-polarized pulses resonant with the $3_{30} \leftrightarrow 2_{21}$ transition (blue arrows in Fig.~\ref{fig:cooling_degenerate}(a)). The population of $\ket{3_{30},1}$ then decays to $\ket{3_{31},1}$ \change{in the cooling step, and the protocol reverts to the first stage of the protocol. Repeating} the complete microwave and cooling sequence several times \change{drives all population into} the target state, \change{as} shown in Fig.~\ref{fig:cooling_degenerate}(b). The target state can be attained with arbitrarily good fidelity for perfectly polarized pulses.
When allowing for non-perfect polarization of the microwave pulses, the final fidelity is sensitive to the imperfection, cf. \change{dashed and dotted lines in} Fig.~\ref{fig:cooling_degenerate}(b). 
The error arises because degenerate states are distinguished only by polarization rather than frequency.
A clean circular polarization of the microwaves~\cite{Yuan_RevSciInstr_2023} is thus essential for \change{preparing a single degenerate quantum state}. 

\change{When designing an experiment, several questions arise. For example, co-trapping the molecule with a single atomic ion is also an option. This will slow down that sideband cooling by a factor of two.
An important question is the choice of rotational state-selective detection, which could be resonance-enhanced multiphoton dissociation~\cite{Staanum_2010, Schneider_2010,Lien_2014} or leverage non-destructive methods such as microwave spectroscopy or quantum logic spectroscopy~\cite{WolfNature_2016,ChouNature2017,Sin22b}.}
One may also wonder whether micromotion of the trapped ions may limit the temperature to which the rotational degrees of freedom can be cooled. This can be alleviated by careful design of the ion trap~\cite{Keller_2019,Kalincev_2021}.
\change{Similarly, Zeeman splitting of the rotational states due to the magnetic field that defines the quantization axis for the laser cooling does not pose an obstacle. Magnetic fields below one Gauss are sufficient~\cite{Kalincev_2021}. Protonated 1,2-propanediol is a closed-shell molecule and does not have a magnetic dipole moment; and to within $1\,$kHz no energy level shifts have been observed for closed-shell molecules in such magnetic fields~\cite{Melanie}. }

Our protocol for cooling the rotational degrees of freedom is easily adapted to \change{other polyatomic} molecular ions. \change{While for protonated 1,2-propanediol},  the protocol relied on the closely spaced pairs of rotational levels of a near prolate asymmetric top, similar level spacings can also be found in near oblate asymmetric tops, \change{and in polyatomic molecules with hyperfine structure, cf. SM~\cite{SM}.
For molecules with non-negligible Zeeman splitting in a small magnetic field, only a single rotational transition will contribute to the resonant dipole-phonon coupling. This might require a larger number of microwave drives, but does not preclude cooling, since a single decay channel is sufficient.} In order to be amenable to rotational sympathetic cooling, the molecular ions have to be polar and of $C_1$ or $C_s$ symmetry, to ensure \change{at least two} non-vanishing Cartesian components of the \change{intrinsic molecular} dipole moment and thus controllability within every rotational subspace~\cite{Leibscher22}. 
The protocol can also be used for diatomic polar molecular ions \change{with $\Lambda$ or $\Omega$ doubling~\cite{Brown_Carrington_2003,Gordon_Field_2025}:} They possess rotational level spacings that can become resonant to \change{trap} normal mode frequencies~\cite{Campbell_PRL_2020}, and microwave control of the rotational dynamics is asserted by complete controllability of linear rotors~\cite{Judson1990}.

In conclusion, we have shown how resonant dipole-phonon coupling enables sympathetic sideband cooling of the rotations of polyatomic molecules. Combining sideband cooling and coherent microwave excitation, an arbitrary subspace of the rotational spectrum of an asymmetric top can be depleted, transferring all population into a single pure rotational state. The cooling scheme requires strong resonant coupling between the rotational degrees of freedom and the collective \change{translational} motion of the trapped particles. A large molecular dipole moment and closely lying rotational levels are thus favorable. 
Asymmetric top molecules possess dipole-allowed rotational transitions that allow for such resonant dipole-phonon coupling -- the cooling scheme is thus particularly well-suited for polyatomic molecules. 
Our protocol enables the preparation of polyatomic molecular ions in a single quantum state. It thus opens the door to extending quantum logic spectroscopy from diatomic to polyatomic molecules and to using the rich rotational spectrum of asymmetric top rotors in fundamental physics experiments or quantum simulation and error correction. 

\begin{acknowledgments}
\change{ We thank Freya E. L. Berggötz for assistance in calculating the rotational constants of protonated 1,2-propanediol and}
gratefully acknowledge financial support from the Deutsche Forschungsgemeinschaft through CRC 1319 ELCH and the joint ANR-DFG CoRoMo project 505622963 (KO 2301/15-1).
\end{acknowledgments}

\bibliography{trap.bib}

\section*{End Matter}

\begin{appendix}
\subsection{Appendix A: Resonant dipole-phonon interaction}
Here, we provide the details of the interaction between the rotation of an asymmetric top molecular ion and a normal mode of the linear Paul trap, as depicted in Fig.~\ref{fig:trap}(a). 
\change{The rotational states are the asymmetric-top eigenstates $\ket{j} = \ket{J_{K_a K_c},M}$, where $K_a K_c = 0,1,\ldots J$  and $M=-J,-J+1,\ldots,J$ with $J=0,1,2,\ldots$ the rotational quantum number. 
}
For simplicity, we omit \change{here} the internal states of the atoms which are not affected by the dipole-phonon interaction.
The collective \change{translational} motion of the charged particles \change{in the trap} is well described by a set of harmonic oscillators, $H_{vib}= \sum_p \hbar \omega_p \left( a_p^\dagger a_p + \frac{1}{2} \right)$ with eigenstates $\ket{n_p}$, $n_p=0,1,2,...$ for each normal mode $p$. The bare states, without \change{the dipole-phonon} interaction, are denoted by $|n_p,j\rangle = \ket{n_p} \otimes \ket{j}$.
The interaction Hamiltonian can thus be expressed as $H_{dp} = \sum_p H_{dp}^{(p)}$, where the interaction of ${\bm\mu}$ with normal mode $p$ is given by
\begin{equation}
H_{dp}^{(p)} = {\cal E}^{(p)}_0 (a_p + a_p^\dagger) \left( R_{\beta a} \mu_a + R_{\beta b} \mu_b + R_{\beta c} \mu_c \right)\,,
\label{eq:dipole_asym_top}
\end{equation}
where indices $\alpha=a,b,c$ and $\beta=x,y,z$ refer to molecule-fixed, resp. space-fixed axes.
The elements of the rotation matrix $R_{ \beta \alpha}=R_{ \beta \alpha}(\gamma_R)$ with $\gamma_R$ denoting the Euler angles 
determine the selection rules~\cite{Zare88} 
with $\beta=x$, resp. $\beta=z$, for interaction with an axial, resp. radial, mode. 
The dipole-phonon interaction becomes strong if one of the normal modes is resonant to a dipole-allowed rotational transition~\cite{HudsonPRA2018,Campbell_PRL_2020}. The interaction with all other normal modes can then be neglected and effectively $H_{dp}=H_{dp}^{(p)}$. In this case, the pair of degenerate states $\ket{n_p,j_1}$ and $\ket{n_p-1,j_2}$ with $n_p >0$ couple, and the degeneracy is lifted by the splitting $\Delta E_{n_p}$ as described in Eq.~(\ref{eq:dipole_splitting}).
For the example of the rotational transition between the states $\ket{3_{31},M}$ and $\ket{3_{30},M}$ of 1,2-propanediol matching the radial zig-zag mode frequency,  $H_{dp} = {\cal E}^{(p)}_0 (a_p + a_p^\dagger) R_{z a} \mu_a$ such that Eq.~\eqref{eq:dipole_splitting} becomes
\begin{equation}
    \Delta E_{n_p} = \sqrt{n_p} \left |{\cal E}^{(p)}_0 \mu_a 
    \bra{3_{30},M} R_{za} \ket{3_{31},M} \right |\,.
    \label{eq:dipole_splitting_2}
\end{equation}
The interaction strength ${\cal E}^{(p)}_0 =  b_m^{(p)} \sqrt{\frac{\hbar}{2e^2} \omega_p^3 M_{rot}}$ 
depends on the normal mode frequency $\omega_p$, the total mass of the molecule $M_{rot}$, and the displacement of the center of mass of the molecule $b_m^{(p)}$\cite{Campbell_PRL_2020}.

\subsection{Appendix B: Sympathetic sideband cooling with dipole-phonon interaction}

To model sympathetic sideband cooling, we consider a laser beam interacting with the atomic ions such that the total Hamiltonian reads
\begin{equation}\label{eq:H_of_t}
  H(t) = H_0 + H_{rot} + H_{dp} + \sum_{i=1}^2 \left( H_a^{(i)} + H_{al}^{(i)} (t) \right)
\end{equation}
with $H_0=\omega_p a_p^\dagger a_p$ describing the normal mode $p$ that is  resonant with a rotational transition. 
For simplicity, we do not account for the remaining normal modes which do not contribute to the dipole-phonon coupling. For a 1,2-propanediol ion, the transition between $\ket{j_1}=\ket{3_{31},M}$ and $\ket{j_2}=\ket{3_{30},M}$ for $-3 \leq M \leq 3$ is resonantly coupled to the radial zig-zag mode. Since the interaction does not couple rotational rotational states with different $M$, sideband cooling can be simulated for each value of $M$ separately.   
Considering only the two rotational states $\ket{j_1}$ and $\ket{j_2}$,  the rotational Hamiltonian is
\begin{equation}
   H_{rot} = E_{j_1} \ket{j_1}\bra{j_1} + E_{j_2} \ket{j_2}\bra{j_2}\,,
\end{equation}
whereas $H_{dp}=H_{dp}^{(p)}$ describing the dipole-phonon interaction for normal mode $p$ is given in Eq.~\eqref{eq:dipole_asym_top}.
Denoting the internal states of the atomic ions $\ket{a_i}$ with  $a_i=g_i,e_i$, $H_a^{(i)}=\omega_a \ket{e_i} \bra{e_i}$ with $\omega_a$ the atomic excitation energy. The atoms interact with a laser beam with frequency $\omega_l$ via 
\begin{equation*}
  H_{al}^{(i)}(t) = \frac{\Omega}{2} \left(\sigma_+^{(i)} + \sigma_-^{(i)}\right) \exp{[i ( {\bf k}\cdot {\bf r}_i - \omega_l t)]} + h.c. \,,
\end{equation*}
where $\Omega$ is the Rabi frequency and ${\bf r}_i$ the position of atom $i$. Considering only one normal mode, namely the zig-zag mode in $z$-direction, this becomes
\begin{equation*}
  H_{al}^{(i)}(t) = \frac{\Omega}{2} \left(\sigma_+^{(i)} + \sigma_-^{(i)}\right) \exp{[i ( k_z z_i - \omega_l t + \phi_i)]} + h.c. 
\end{equation*}
with $\phi_i = k_x x_{0i}$ the phase of the electric field at the equilibrium position $x_{0i}$ of atom $i$, cf. Fig.~\ref{fig:trap}.
We assume the Lamb-Dicke limit \cite{Leibfried_2003}, where
$\exp{[\pm i k_z z_i]} \approx 1 \pm i k_z z_i = 1 \pm i \eta \, b_i^{(p)} (a_p + a_p^\dagger)$ with the Lamb-Dicke parameter $\eta = \sqrt{\hbar/(2 M_a \omega_p)}$ and $b_{i=1,2}^{(p)}$ the (normalized) displacement of the two atoms for normal mode $p$.
Going into the interaction picture with respect to $H_0+H_{rot}+H_{a}$
and invoking the rotating wave approximation results in
\begin{eqnarray}
    H_{al,I}^{(i)}(t) &=& \frac{\Omega}{2} e^{i (\Delta t+\phi_i)} \sigma_+ 
    \left[ 1 + i \eta \, b_i^{(p)} \left(e^{i \omega_p t} a + e^{-i \omega_p t} a^\dagger\right) \right] \nonumber \\
     &&+ h.c.\,,
\end{eqnarray}
where $\Delta = \omega_a - \omega_l$ and, for Eq.~\eqref{eq:dipole_asym_top}, in 
\begin{equation}
H_{dp,I} = {\tilde {\cal E}}_0^{(p)} \mu_a \left( a_p \ket{j_2}\bra{j_1} + a_p^\dagger \ket{j_1}\bra{j_2} \right)
\end{equation}
with ${\tilde {\cal E}}_0^{(p)} = {\cal E}_0^{(p)}
\bra{3_{30},M}R_{za}\ket{3_{31},M}$ depending on the orientational quantum number $M$.

The master equation capturing coherent dynamics under laser excitation and the spontaneous emission of the atomic ions 
reads
\begin{eqnarray}
    \frac{\partial \rho}{\partial t} = - \frac{i}{\hbar} \left[ H_I(t), \rho \right] +
    {\cal L}_D \rho\,,
    \label{eq:Lindblad}
\end{eqnarray}
where $H_I(t)=\sum_{i=1,2} H_{al,I}^{(i)}(t) +H_{dp,I}$
and ${\cal L}_D \rho = L \rho L^\dagger - \frac{1}{2} \left[ L^\dagger L, \rho \right]$\cite{BreuerBook}.
We assume that the spontaneous decay is not affected by the interactions, i.e., we use the standard jump operator $L = \gamma\sum_{i=1}^{2}  \sigma_-^{(i)}$ and only a local master equation~\cite{LevyEPL2014}. Given the huge separation in transition energies, this is well justified here.

We numerically solve Eq.~\eqref{eq:Lindblad} using the QuTiP package \cite{johansson2012qutip}. We consider a Hilbert space spanned by the atomic states $\ket{a_i} = \ket{g_i/e_i}$ for $i=1,2$, the \change{phonon} states $\ket{n_p} =\ket{0,1,2}$ and the rotational states $\ket{j_1}$, $\ket{j_2}$. For a protonated 1,2-propanediol, the rotational constants are $A=8612\,$MHz, $B=3758\,$MHz and $C=2868\,$MHz \cite{SM}. For sideband cooling, the laser pulse is red-shifted, i.e., $\omega_l = \omega_a - \omega_p$.
The values of the Rabi frequency, Lamb-Dicke parameter and decay rate for sideband cooling of Yb$^+$ ions, see Fig.~\ref{fig:sideband_cooling}, are taken from \cite{Kulosa_2023}. Further details are given in the SM \cite{SM}. Initially, the atomic ions are taken to be in their electronic ground state, and, for simplicity, an equal distribution of population in the states $\{\ket{n_p,j}\}$, $n_p=0,1,2$ and $j=j_1,j_2$ is assumed. 
\end{appendix}

\end{document}


\title{Supplemental Material\\ Sympathetic rotational cooling of large trapped molecular ions}
\author{Monika Leibscher}
\affiliation{Freie Universit\"{a}t Berlin, Dahlem Center for Complex Quantum Systems and Fachbereich Physik, Arnimallee 14, 14195 Berlin, Germany}
\author{Alexander Blech}
\affiliation{Freie Universit\"{a}t Berlin, Dahlem Center for Complex Quantum Systems and Fachbereich Physik, Arnimallee 14, 14195 Berlin, Germany}
\author{Christiane P. Koch} 
\email{Email: christiane.koch@fu-berlin.de}
\affiliation{Freie Universit\"{a}t Berlin, Dahlem Center for Complex Quantum Systems and Fachbereich Physik, Arnimallee 14, 14195 Berlin, Germany}
\date{\today} 

\maketitle


We document here the details of the numerical calculations
and show results for additional examples, in support of our claim that resonant dipole-phonon coupling and sympathetic sideband cooling for rotational degrees of freedom is feasible for a large range of polyatomic molecules.

\section{Dipole-phonon coupling}
The strength of the dipole-phonon interaction for normal mode $p$, cf. Eq.~(10) in the main text, is given by
\begin{equation}
{\tilde {\cal E}}_0^{(p)} = {\cal E}_0^{(p)}
\mu_a \bra{J_{K_a,K_c},M}R_{za}\ket{J_{K_a,K_c-1},M}  \nonumber
\end{equation}
for a near prolate asymmetric top.
The prefactor
\begin{equation}
    {\cal E}^{(p)}_0 =  b_m^{(p)} \sqrt{\frac{\hbar}{2e^2} \omega_p^3 M_{rot}}
    \nonumber
\end{equation}
depends on the mass of the molecule $M_{rot}$, the normal mode frequency $\omega_p$ and the displacement of the center of mass of the molecule $b_m^{(p)}$ \cite{Campbell_PRL_2020}. The latter two, $\omega_p$
and $b_m^{(p)}$, in turn depend on the axial and radial trap frequencies $\omega_x$ and $\omega_z$ \cite{James_ApplPhys_1998,James_PRL_2000}.  
Strong dipole-phonon interaction thus requires a large $\omega_p$. 

\subsection{Choice of the normal mode for resonant dipole-phonon coupling}
Since radial normal modes typically have a higher frequency than the axial normal modes, it is favorable to tune the trap such that one of the radial modes is in resonance with a rotational transition. In a linear trap with three ions, the radial modes with $b_m^{(p)}\neq 0$ are the center-of-mass and zig-zag modes. Both modes have comparable frequencies, but for the zig-zag mode, $|b_m^{(p)}|$ is larger. The dipole-phonon coupling is thus particularly strong for coupling to the radial zig-zag mode. Therefore, we only consider resonant coupling to this mode.  

\subsection{Dependence on molecular parameters}
To demonstrate that conditions for resonant dipole-phonon interaction can be fulfilled for a large range of polyatomic molecular ions, we consider examples with molecular mass ranging from about 70$\,$u to 260$\,$u. For significantly lighter molecules, the dipole–phonon coupling will be too weak, whereas it is increasingly challenging to realize sufficiently high trap frequencies for heavier species. A list of example molecular ions with relevant molecular parameters is given in 
Table~\ref{tab:rotational_constants}.
\begin{table}[tbp]
    \centering
    \begin{tabular}{c|ccccc}
         & $M_{rot}$ & $A$ & $B$ & $C$ & $\mu_a$\\
         \hline \\
        H-propanediol$^+$  & 77 & 8612 & 3758 & 2868 &
         2.4 \\ 
        H-glutamine$^+$ {\cite{Pang_2013}}  & 147 & 2210 & 830   & 750 & 2.0 \\
        CHDBrI$^+$ {\cite{Landau_2023}} & 221 &  14750 & 1170 & 1095\\
        CHCaBrI$^+$ {\cite{Landau_2023}} & 259 & 2550 &830  & 660\\    
    \end{tabular}
    \caption{Molecular mass $M_{rot}$ is given in u, rotational constants $A$, $B$, and $C$ in MHz and the dipole moment $\mu_a$ in D.}
    \label{tab:rotational_constants}
\end{table}
The parameters for protonated glutamine \cite{Pang_2013}, CHDBrI$^+$ and CHCaBrI$^+$ \cite{Landau_2023} are found in the literature, whereas for protonated 1,2-propanediol, we have carried out quantum chemical calculations with details reported below.

Protonated propanediol and glutamine are examples for small and medium sized organic molecular ions that could be candidates for quantum error correction, whereas  CHDBrI$^+$ and CHCaBrI$^+$ are considered as candidates for measuring parity violation \cite{Landau_2023}. All examples are near prolate asymmetric tops, i.e., the resonant rotational transitions are a-type transitions between the states $J_{K_a,K_c}$ and $J_{K_a,K_{c+1}}$, and the relevant component of the molecular dipole moment is $\mu_a$. 
The conditions for resonant dipole-phonon coupling can also be fulfilled for near oblate asymmetric tops such as the astrophysically relevant protonated oxirane \cite{Puzzarini_2014}, where the resonant rotational transitions are of c-type and occur between states $J_{K_a,K_c}$ and $J_{K_{a+1},K_c}$. As such, they allow for dipole-phonon coupling if the component $\mu_c$ of the molecular dipole moment is non-zero. 
In fact, as long as the molecule is an asymmetric top in the mass range above and has $C_1$ or $C_s$ symmetry (the latter is necessary to ensure that at least two components of the molecular dipole moment are non-zero), it is very likely to find a dipole-allowed rotational transition of the order of a few MHz that allows for resonant dipole-phonon coupling.

\begin{figure}[tbp]
    \includegraphics[width=1.0\linewidth]{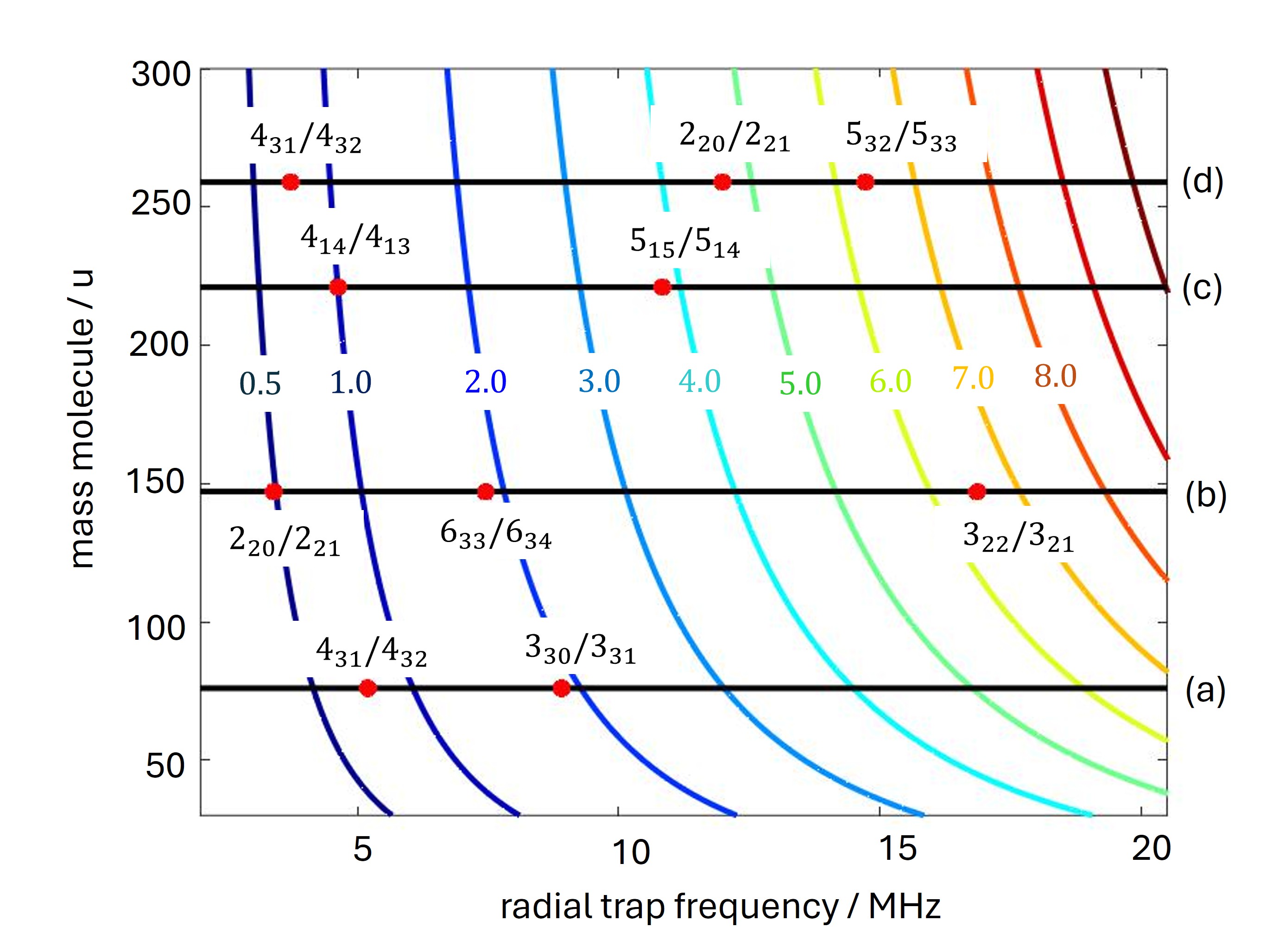}
	\caption{Dipole-phonon coupling strength $|{\cal E}^{(p)}_0|$ (contour lines) in units of kHz/D as function of radial trap frequency and molecular mass. The axial trap frequency is $\omega_x=1$ MHz. The horizontal black lines indicate the mass of protonated propanediol (a), protonated glutamine (b), CHDBrI$^+$ (c), and CHCaBrI$^+$ (d). Dipole allowed transitions between the rotational states
    $J_{K_a,K_c}$ and $J_{K_a,K_{c+1}}$ which are resonant to the frequency $\omega_p$ of the radial zig-zag mode are denoted by red stars. We consider a linear trap, where the molecular ion is trapped between two $^{172}$Yb$^+$ ions.}
	\label{fig:coupling_mass_frequency}
\end{figure}
For the molecules listed in Table~\ref{tab:rotational_constants}, 
Fig.~\ref{fig:coupling_mass_frequency} shows the dipole-phonon coupling strength  $|{\cal E}^{(p)}_0|$ for the radial zig-zag mode as a function of the radial trap frequency and the mass of the molecular ion. 
Resonant dipole-phonon coupling requires that the energy difference between two  rotational states (connected by an electric dipole allowed transition) is equal to the normal mode frequency $\omega_p$, here the radial zig-zag mode. The red stars in Fig.~\ref{fig:coupling_mass_frequency} mark transitions and radial trap frequencies, for which the resonance condition is fulfilled. 
For all example molecules, we find rotational transitions that fulfill the resonance conditions for radial trap frequencies smaller than $20$ MHz, with coupling strengths  $|{\cal E}^{(p)}_0|$ of the order of kHz/D. To obtain the total interaction strength ${\tilde {\cal E}}_0^{(p)}$, ${\cal E}_0^{(p)}$ is multiplied by the molecular dipole moment $\mu_a$ in Debye and the transition matrix element
$\bra{J_{K_a,K_c},M}R_{za}\ket{J_{K_a,K_c-1},M}$ for the resonant rotational transition, which depends on the orientational quantum number $M$. Note that $1 > |\bra{J_{K_a,K_c},M}R_{za}\ket{J_{K_a,K_c-1},M}| \ge 0$. For $M=0$ the transition matrix element is zero, and for a given $J$ it is largest for $M=J$. 
While the results for molecules with strong hyperfine interaction are only indicative, cf. Sec.~\ref{sec:hyperfine} below, Fig.~\ref{fig:coupling_mass_frequency} nevertheless suggests that suitable rotational transitions with resonance conditions at feasible trap frequencies can be found for a large variety of molecules.

\subsection{Quantum chemical calculations}
For protonated 1,2-propanediol, we have calculated molecular constants using the ORCA quantum chemistry package~\cite{ORCA6.0}.
A protonation-site scan using CREST (Conformer-Rotamer Ensemble Sampling Tool)~\cite{Prach_2020} identified the two OH groups as possible protonation sites. For both protonated species, we have carried out conformational searches using the extended tight-binding method \cite{Bannwarth_2019,deSouza_2025} and optimized all unique minima using second-order M\o{}ller–Plesset perturbation theory \cite{MollerPlesset_1934} with the aug-cc-pVTZ basis set \cite{Dunning_1989,Kendall_1992}. This procedure yielded two conformers for protonation at the primary OH group and seven for protonation at the secondary OH group.
We have selected the lowest-energy conformer of the former, as it has the largest dipole moment component along the $a$-axis. The molecular constants listed in Table~\ref{tab:rotational_constants} refer to this conformer.

\section{Rotational cooling via sympathetic sideband cooling}
Here, we present further examples as well as the numerical details for the simulation of sympathetic sideband cooling of rotational states {\it via} resonant dipole-phonon coupling.
Sideband cooling relies on the excitation of the atomic ions with a red-shifted laser beam. The interaction between atom $i=1,2$ with the laser beam in the Lamb-Dicke limit is given by (see Eq.(9) in the main text)
\begin{eqnarray}
    H_{al,I}^{(i)}(t) &=& \frac{\Omega}{2} e^{i (\Delta t+\phi_i)} \sigma_+ 
    \left[ 1 + i {\tilde \eta} \left(e^{i \omega_p t} a + e^{-i \omega_p t} a^\dagger\right) \right] \nonumber \\
     &&+ h.c.\,,
     \nonumber
\end{eqnarray}
with $\tilde{\eta}=\eta \, b_i^{(p)}$, and $\eta=k x_0$ with wavenumber $k$ and $x_0=\sqrt{\hbar/(2 M_a \omega_p)}$ being the Lamb-Dicke parameter. Here, $M_a$ is the mass of the atomic ions and $b_i^{(p)}$ is their displacement for normal mode $p$, here the radial zig-zag mode. For sideband cooling, the laser is tuned to the red sideband, i.e. $\Delta = \omega_0 - \omega_L = -\omega_p$.
For $^{172}\mbox{Yb}^+$, sideband cooling is carried out on the
electronic $^2S_{1/2} \rightarrow \, ^2D_{5/2}$ transition (411 nm). The $^2D_{5/2}$ state is coupled to the short-lived $^2P_{3/2}$ state by a laser beam with wavelength of 1650 nm. 
The decay rates of the atomic states are given in \cite{Kulosa_2023}. For the atomic ions, Raman sideband cooling can be described by an effective two-state model \cite{Marzoli1994} with effective wave vector ${\vec k}= \vec{k}_1 - \vec{k}_2$, where $\vec{k}_1$ and $\vec{k}_2$ are the wave vectors of the two Raman beams. It depends on the two Rabi frequencies for the two Raman laser pulses that can be adjusted. In the effective two-state model, this results in the effective decay rate $\gamma_{eff}$ and the Raman frequency $\Omega$ that can be controlled in order to optimize the cooling \cite{Kulosa_2023}. 
We assume that the translational motion of the trapped particles is cooled to the lowest few vibrational states. Possible thermalization of the particle due to black body radiation is not considered here since it takes place on a time of seconds while the cooling process is much faster\cite{Vilas_PRA_2023}.

\begin{figure*}[tbp]
\includegraphics[width=0.9\linewidth]{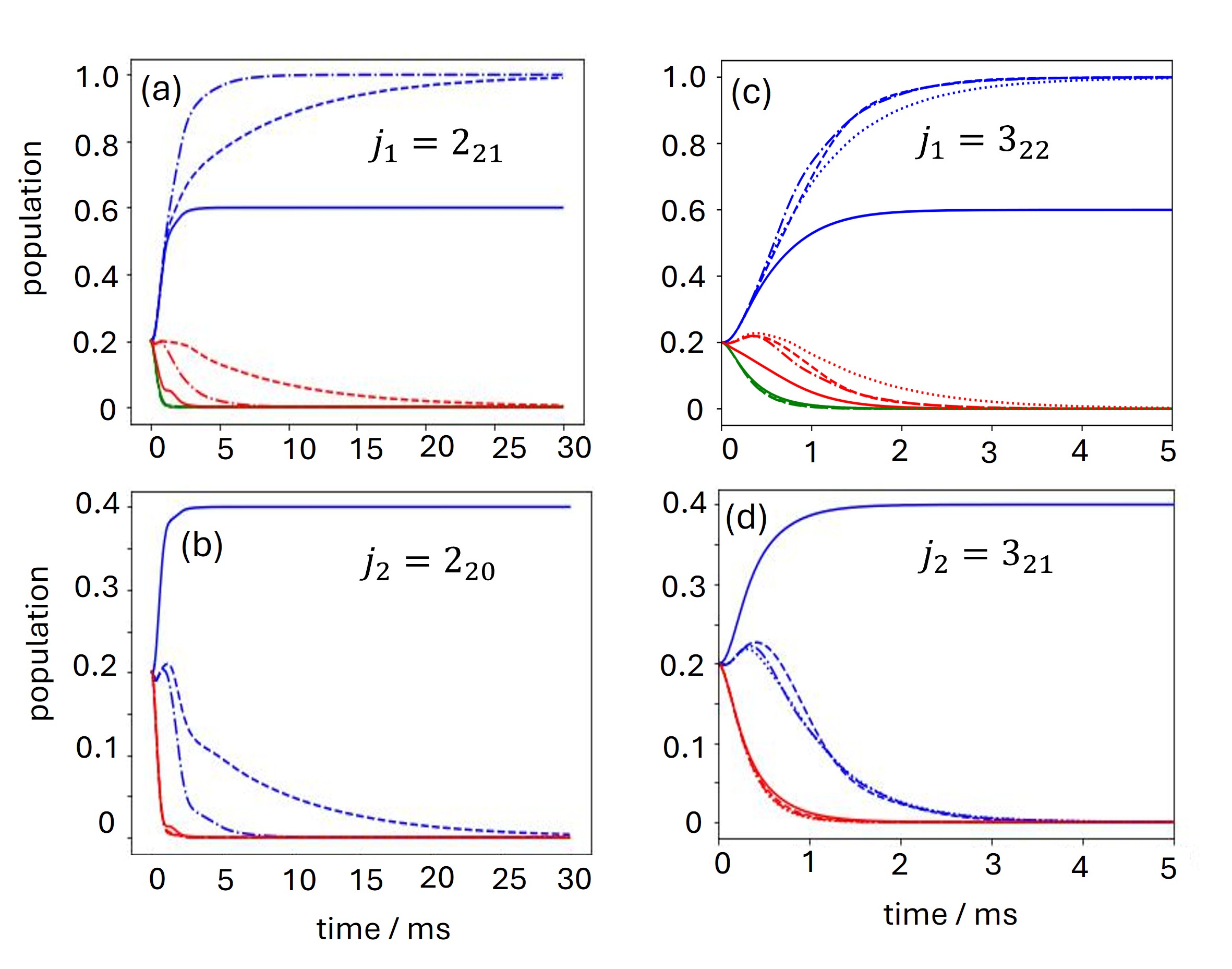}
	\caption{Rotational sideband-cooling for glutamine co-trapped with two $^{172}$Yb$^+$ ions.
    (a,c): Population of the lower rotational states $j_1=2_{21}$ and $j_1=3_{22}$.
    (b,d): Population in the higher rotational states$j_2=2_{20}$ and $j_1=3_{21}$. 
    Blue, red and green lines depict the phonon states ($n_p=0,1,2$); solid, dashed, dash-dotted and dotted lines refer to $M=0,1,2$ and $3$. All populations are for the electronic ground state of the atoms.  The parameters are given in Table \ref{tab:cooling_parameter}.}
	\label{fig:cooling_examples}
\end{figure*}
\begin{table}[tbp]
    \centering
    \begin{tabular}{c|ccc}
         & H-glutamine & H-glutamine & H-propanediol \\
         & $2_{21}/2_{20}$  & $3_{22}/3_{21}$ & $3_{31}/3_{30}$\\
         \hline \\
     $\omega_x$/MHz    & 1.0 & 1.0 &  1.0\\
     $\omega_z$/MHz    & 3.76 & 16.95 &  8.87\\
     $ \tilde{\eta} $    & 0.033 & 0.015 & 0.012\\
     $\tilde{{\cal E}}^{(p)}_0$/kHz &  &  &   \\ 
                   $M=1$         & 0.38 & 2.2  &  2.7 \\ 
                   $M=2$         & 0.8 & 4.3  &  5.7  \\
                   $M=3$         &  & 6.6     &  8.4  \\
     $\gamma_{eff}$ / MHz    & 0.05 & 0.1 & 0.1\\
     $\Omega$ / MHz   & 0.05 & 0.2 & 0.2 \\
    \end{tabular}
    \caption{Parameter for rotational sideband cooling of protonated glutamine and propanediol molecules.}
    \label{tab:cooling_parameter}
\end{table}
To simulate the sideband cooling, we numerically solve the master equation, Eq.(4) in the main text,  using the QuTiP package \cite{johansson2012qutip}. We consider a Hilbert space spanned by the atomic states $\ket{a_i} = \ket{g_i/e_i}$ for $i=1,2$, the phonon states $\ket{n_p} =\ket{0,1,2}$ and the rotational states $\ket{j_1}$, $\ket{j_2}$. The results of our simulation of sympathetic rotational cooling for H-propanediol are shown in the main text. In analogy with Fig.~3 in the main text, Fig.~\ref{fig:cooling_examples} displays the results for H-glutamine, assuming that the trap is adjusted such that either the rotational transitions $2_{21}/2_{20}$ (a,b) or $3_{22}/3_{21}$ (c,d) are resonant with the normal mode frequency $\omega_p$. As expected from Fig.~\ref{fig:coupling_mass_frequency}, the larger coupling strength of the  $3_{22}/3_{21}$ transition results in faster cooling. Indeed, a dipole-phonon coupling strength of a few kHz is sufficient to achieve cooling within less then 10 ms.
It requires, however, a stiff trap with large radial trap frequency ($\omega_z=8.9\,$MHz for propanediol and $\omega_z=17\,$MHz for glutamine). For the $2_{21}/2_{20}$ transition of glutamine, a radial trap frequency of $\omega_z=3.8$ MHz is sufficient to ensure the resonance condition, but the coupling strength is smaller, namely less then a kHz. Rotational cooling can still be achieved, but requires somewhat longer times, about 30 ms for the chosen effective decay rate $\gamma_{eff}$ and Rabi frequency $\Omega$, well within in the range considered in Ref.~\cite{Kulosa_2023}. Optimization of the protocol parameters is likely to speed up the cooling process.

For the examples in Fig.~3 of the main text and Fig.~\ref{fig:cooling_examples}, 
we have considered rotational transitions with the lowest possible $J$. This simplifies the microwave excitation that is interleaved with the sideband cooling. However, the controllability of asymmetric top rotors~\cite{Leibscher22,Pozzoli21} implies the existence of microwave excitation schemes for arbitrary rotational transitions and thus guarantees rotational cooling with a combination of sideband cooling and microwave excitation, irrespective of the specific choice of $J$.

The molecular ions CHDBrI$^+$ and CHCaBrI$^+$ also display resonant dipole-phonon coupling in the range of kHz and are therefore also amenable to sympathetic rotational cooling. Since these molecules have a large nuclear quadrupole moment, the coupling between the nuclear spin and orbital angular momenta have to be considered in the rotational spectrum, as discussed below. A detailed investigation of their rotational cooling is therefore left to a future study.

\section{Role of hyperfine interaction}
\label{sec:hyperfine}
In order to focus on the key points for sympathetic rotational cooling, our model assumes a pure asymmetric top rotational spectrum, neglecting hyperfine interaction.
Protonated propanediol does not contain nuclei with quadrupole moment and can thus be reasonably well described as an asymmetric top molecule. By contrast, H-glutamine, CHDBrI$^+$ and CHCaBrI$^+$ have one or more nuclei with a spin $I \ge 1$ and thus a non-vanishing nuclear quadrupole moment. 
For an exact calculation of the rotational energies and the conditions for resonant dipole-phonon coupling, the hyperfine splitting of the rotational states has to be taken into account. For the sake of simplicity, we have approximated the rotational structure as that of an asymmetric top without any spin-orbit couplings. 

When designing an experiment, the full spin-rotation structure will need to be considered. Importantly, adding the hyperfine interaction does not impede the resonant dipole-phonon coupling~\cite{Campbell_PRL_2020} on which our protocol relies. However, the molecular state space gets larger. While this will complicate the practical implementation of the protocol, the presence of dipole-allowed transitions between hyperfine states allows for population flow to the cooled rotational level. Two strategies can be employed to this end: For smaller splittings, as expected for protonated glutamine, chirping the microwave pulses allows to  drive population transfer in near-degenerate transitions simultaneously, as demonstrated recently in neutral valinol molecules~\cite{BerggoetzJPCL25}. For large splittings, the number of microwave drives needs to be increased but the different transitions can be addressed sequentially.

\bibliography{trap.bib}